# Influence of the Range of Interaction on the Stability of the Nucleus propagating with Momentum Dependent Interactions

Supriya Goyal[1], Rajeev K. Puri[1]
[1]Department of Physics, Panjab University, Chandigarh-160014, India.

The behavior of the hot and dense nuclear matter at extreme conditions of temperature and density can be studied with the help of heavy-ion collisions at intermediate energies. The model used for the present study is the Quantum Molecular Dynamics (QMD) model which treats every nucleon as Gaussian wave packet [1]. The role of the range of interaction of Gaussian wave packets on the stability of the nuclei propagating with momentum dependent interactions is analyzed. Comparison is done by using the standard (1.08 fm$^2$) and double (2.16 fm$^2$) width of the Gaussian wave packets for a large number of nuclei over the entire periodic table.

The role of the momentum dependent interactions can not be neglected in a realistic treatment of heavy-ion collisions [2, 3]. These interactions are repulsive in nature and thus the nuclei generated with these interactions are not stable even in the initial stage [2, 3]. We find that nuclei having double width of wave packets do not emit free nucleons for a long period of time. The ground state properties of all the nuclei are also described well. For clarity, the trajectories of the 13 free nucleons (in the x-z plane) emitted from the $^{197}$Au nucleus having standard (left panel) and double (right panel) width is shown in Fig 1. It is clear from the figure that the role of the double width can not be neglected while initializing the nuclei with QMD model.

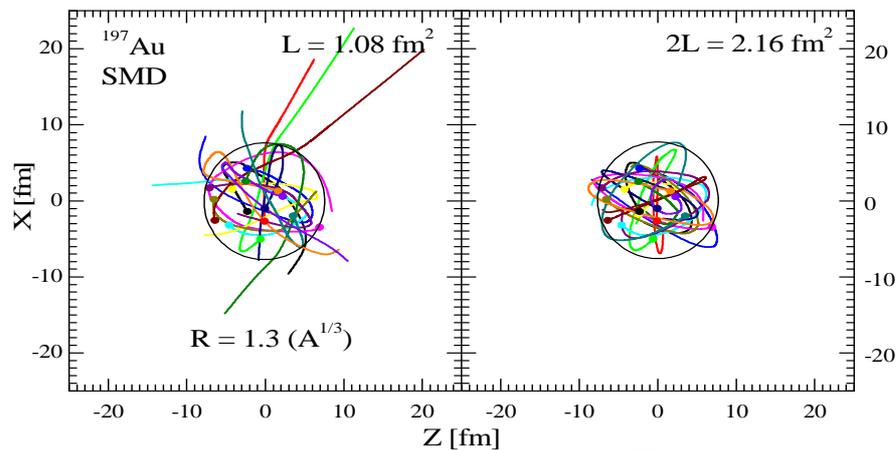

Figure 1: The trajectories of the free nucleons emitted from the $^{197}$Au nucleus with standard (left panel) and double (right panel) width of the Gaussian wave packets in the x-z plane. The solid symbols show the position of the nucleons at 0.1 fm/c.